\documentclass{appolb}
\usepackage{graphicx}
\usepackage{amsmath}
\usepackage{mathtools}
\usepackage{subcaption}
\usepackage{placeins}

\renewcommand{\runhead}{}

\begin{document}

\title{Meson production in proton-proton scattering within an extended
linear sigma model\thanks{%
Presented at Excited QCD 2014, Bjelasnica Mountain, Sarajevo}}
\author{Khaled Teilab, Francesco Giacosa, and Dirk H. Rischke %
\address{Institut f\"ur Theoretische Physik, Goethe-Universit\"at Frankfurt} 
}
\maketitle

\begin{abstract}
We study the production of non-strange mesons in proton-proton scattering
reactions within an effective model of mesons and baryons with global chiral
symmetry. The model includes the nucleon ($N$) and its chiral partner ($%
N^{\ast }$) and their couplings to the $N_{f}=2$ multiplets of
(pseudo)scalar and (axial-)vector mesons. Results for the production of $%
\omega $, $\eta $ mesons are presented.
\end{abstract}



\section{Introduction}

Phenomenological models of QCD are suitable for the study of hadrons and
their interactions at energies of a few GeV, where perturbative methods are
not applicable, see e.g. \cite{geffen,bwlee,meissner,kk,mosel,mirror}.

In this work we use an effective chiral model of QCD, denoted as the
extended Linear Sigma Model (eLSM) \cite%
{susanna1,denis2,denis3,susanna2,otherselsm}, in order to study
proton-proton scattering in which a meson is produced in the final state. In
the eLSM, the nucleon ($N$) and its chiral partner ($N^{\ast }$) are
introduced according to the mirror assignment \cite{mirror}. To this end, we
first introduce the baryonic fields $\Psi _{1}$ and $\Psi _{2}$ which
transform under chiral transformations $U(2)_{L}\times U(2)_{R}$ as follows: 
\begin{align}
\Psi _{1R}& \longrightarrow U_{R}\Psi _{1R}\quad ,\quad \Psi
_{1L}\longrightarrow U_{L}\Psi _{1L}\quad ,  \notag \\
\Psi _{2R}& \longrightarrow U_{L}\Psi _{2R}\quad ,\quad \Psi
_{2L}\longrightarrow U_{R}\Psi _{2L}\quad ,
\end{align}%
where $\Psi _{iR}$ and $\Psi _{iL}$ denote the right- and left-handed
components of the baryon field $\Psi _{i}$, respectively. This configuration
allows to write down a chirally symmetric mass term of the baryons of the
form: 
\begin{equation}
\mathcal{L}_{1}=-m_{0}\left( \overline{\Psi }_{1R}\Psi _{2L}+\overline{\Psi }%
_{2L}\Psi _{1R}-\overline{\Psi }_{1L}\Psi _{2R}-\overline{\Psi }_{2R}\Psi
_{1L}\right) \quad .
\end{equation}%
The physical fields 
($N$) and ($N^{\ast }$)
are given by: 
\begin{equation}
\begin{pmatrix}
N \\ 
N^{\ast }%
\end{pmatrix}%
=\frac{1}{\sqrt{2\cosh \delta }}%
\begin{pmatrix}
e^{\delta /2} & \gamma _{5}e^{-\delta /2} \\ 
\gamma _{5}e^{-\delta /2} & -e^{\delta /2}%
\end{pmatrix}%
\begin{pmatrix}
\Psi _{1} \\ 
\Psi _{2}%
\end{pmatrix}%
\quad .
\end{equation}%
The interaction of the baryons with the (pseudo)scalar fields is given by
the term $\mathcal{L}_{2}$: 
\begin{equation}
\mathcal{L}_{2}=-\hat{g}_{1}\left( \overline{\Psi }_{1L}\Phi \Psi _{1R}+%
\overline{\Psi }_{1R}\Phi ^{\dagger }\Psi _{1L}\right) -\hat{g}_{2}\left( 
\overline{\Psi }_{2L}\Phi \Psi _{2R}+\overline{\Psi }_{2R}\Phi ^{\dagger
}\Psi _{2L}\right)
\end{equation}%
where $\Phi $ represents the matrix of states with zero spin: 
\begin{equation}
\Phi =\left( \sigma +i\eta _{N}\right) t^{0}+\left( \boldsymbol{a}_{0}+i%
\boldsymbol{\pi }\right) \cdot \boldsymbol{t}\quad ,
\end{equation}%
with $t^{0}$ and $\boldsymbol{t}$ as the identity and Pauli matrices divided
by 2 respectively and $\eta _{N}$ as the non-strange component of the $\eta $
meson. The interaction with the (axial-)vector states is obtained via the
term $\mathcal{L}_{3}$: 
\begin{equation}
\mathcal{L}_{3}=\overline{\Psi }_{1L}i\gamma _{\mu }D_{1L}^{\mu }\Psi _{1L}+%
\overline{\Psi }_{1R}i\gamma _{\mu }D_{1R}^{\mu }\Psi _{1R}+\overline{\Psi }%
_{2L}i\gamma _{\mu }D_{2R}^{\mu }\Psi _{2L}+\overline{\Psi }_{2R}i\gamma
_{\mu }D_{2L}^{\mu }\Psi _{2R}
\end{equation}%
with: 
\begin{align}
D_{1R}^{\mu }& =\partial ^{\mu }-ic_{1}R^{\mu }\quad ,\quad D_{1L}^{\mu
}=\partial ^{\mu }-ic_{1}L^{\mu }\quad ,  \notag \\
D_{2R}^{\mu }& =\partial ^{\mu }-ic_{2}R^{\mu }\quad ,\quad D_{2L}^{\mu
}=\partial ^{\mu }-ic_{2}L^{\mu }\quad ,
\end{align}%
and 
\begin{align}
R^{\mu }& =\left( \omega ^{\mu }-f_{1}^{\mu }\right) t^{0}+\left( 
\boldsymbol{\rho }^{\mu }-\boldsymbol{a}_{1}^{\mu }\right) \cdot \boldsymbol{%
t}\quad ,  \notag \\
L^{\mu }& =\left( \omega ^{\mu }+f_{1}^{\mu }\right) t^{0}+\left( 
\boldsymbol{\rho }^{\mu }+\boldsymbol{a}_{1}^{\mu }\right) \cdot \boldsymbol{%
t}\quad .
\end{align}

The values of the parameters $m_{0}$, $\hat{g}_{1}$, $\hat{g}_{2}$, $c_{1}$
and $c_{2}$ are fixed using the following physical quantities: the masses of
the nucleon and nucleonic resonance, the decay width $N^{\ast }\rightarrow
N\pi $, the axial coupling constant of the nucleon and the axial coupling
constant of the nucleonic resonance. Further details on the determination of
parameters and assignment of the fields to physical particles are presented
in Refs. \cite{susanna1,denis2,denis3,susanna2}.

\section{Feynman diagrams}

\begin{figure}[tbp]
\centering
\begin{subfigure}{.5\textwidth}
  \centering
  \includegraphics[width=.8\linewidth]{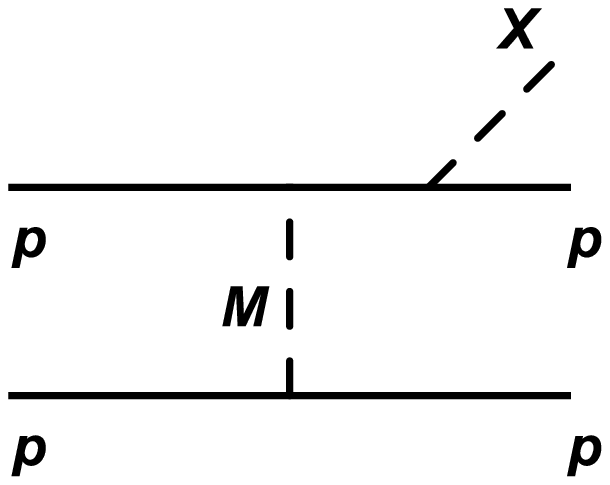}
   \caption{}
  \label{fig:sub1}
\end{subfigure}%
\begin{subfigure}{.5\textwidth}
  \centering
  \includegraphics[width=.8\linewidth]{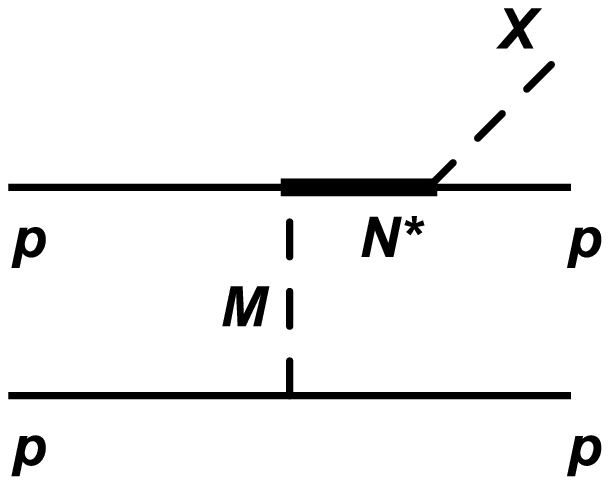}
   \caption{}
  \label{fig:sub2}
\end{subfigure}
\caption{Feynman diagrams of the process $pp\longrightarrow ppX$: a) the
nucleonic current and b) the resonant current. M represents an exchanged $%
\protect\sigma$, $a_0$, $\protect\pi$, $\protect\eta$, $\protect\eta\prime$, 
$\protect\omega$, $\protect\rho$, $f_1$ or $a_1$ meson.}
\label{fig:feyn}
\end{figure}

Using the model described above, we study processes of the type %
\mbox{$pp\longrightarrow ppX$} where $X$ is a single $\omega $ or $\eta $
meson. In figure \ref{fig:feyn}, two sample diagrams of the process are
presented. The diagrams represent either the nucleonic current (fig. \ref%
{fig:sub1}) or the resonant current (fig. \ref{fig:sub2}). To each current
36 different diagrams contribute (9 exchange mesons, $t-$ and $u-$channels
and early and late emission of the produced meson).

The results presented in section \ref{sec:res} are obtained using either
only the nucleonic current or the coherent sum of both the nucleonic and
resonant currents. The corresponding curves are labelled `nucleonic' or
`resonant' respectively.

\section{Results}\label{sec:res} 

\begin{figure}[tbh]
\centerline{\includegraphics[width=0.8\textwidth]{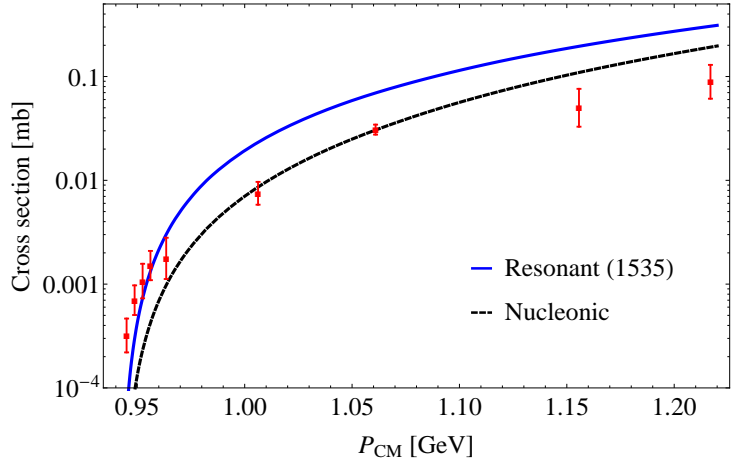}}
\caption{The total cross section for the reaction $pp\longrightarrow pp%
\protect\omega $. The data points show experimental data and are referenced
in \protect\cite{balestraomg}.}
\label{fig:omg}
\end{figure}

The total cross section for the reaction $pp\longrightarrow
pp\omega$ is shown in \mbox{figure \ref{fig:omg}} as a function of the
center-of-mass momentum of the incoming protons. The (black) dashed curve
shows the calculations using only the nucleonic current and the (blue) solid one
shows the calculations using the resonant and nucleonic currents. As seen in
figure \ref{fig:omg}, calculations with the nucleonic current only
underestimate the data points near the production threshold. Including the
resonant contribution leads to a significant improvement of the description
of the data near threshold. It is worth mentioning that the data points
shown in the figure were not used for the determination of the model
parameters. At higher energies, both calculations (nucleonic and resonant)
tend to overestimate the data points.

In the case of the $\eta$ meson, the difference between the nucleonic and
resonant calculations is not significant as can be seen in figure \ref%
{fig:eta}. Again, the (black) dashed curve shows the calculated cross
section considering the nucleonic current only and the (blue) solid curve
shows the sum of the resonant and nucleonic currents.

Both curves overestimate the data
considerably even at threshold.

\begin{figure}[tbh]
\centerline{\includegraphics[width=0.8\textwidth]{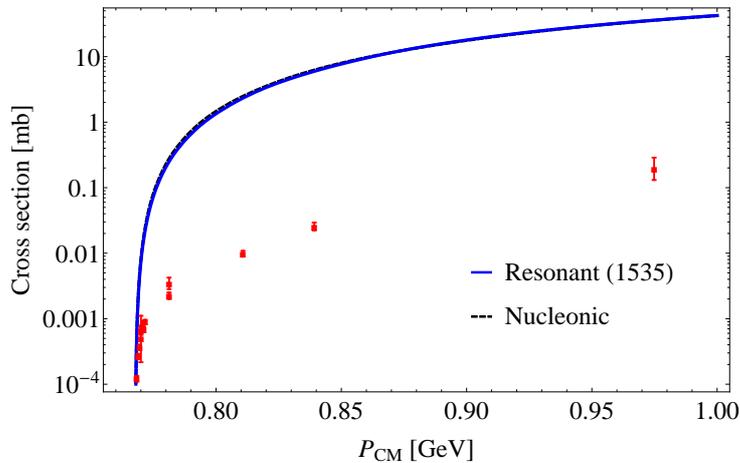}}
\caption{The total cross section for the reaction $pp\longrightarrow pp%
\protect\eta $. The data points show experimental data and are referenced in 
\protect\cite{balestraeta}.}
\label{fig:eta}
\end{figure}

\section{Discussion and Outlook}

The overestimation of the production cross section at high energy
(especially the case of the $\eta $ meson) hints at a missing ingredient in
the calculation. So far we have only considered the production from a
baryon-baryon-meson vertex. The possibility for the produced meson to come
from a three-meson vertex was not included. Including such vertices is
expected to have a considerable effect on the results. In addition, vertex
form factors might be needed for a better agreement with the data at higher
energies.

\end{document}